\pdfoutput=1

\newcommand{\degr}{$^\circ$}
\newcommand{\caproman}[1]{\uppercase\expandafter{\romannumeral#1}}

\newcommand{\BGmail}{bernardgottschalk\,@\,gmail.com}

\newcommand{\eqn}[1]{Eq.\,(\ref{eqn:#1})}

\newcommand{\fig}[1]{Fig.\,\ref{fig:#1}}

\documentclass{article}

\title{\bf Global surface temperature trends\\and the effect of World War \caproman{2}}
\author{Bernard Gottschalk\thanks{Harvard University Laboratory for Particle Physics and Cosmology, 18 Hammond St., Cambridge, MA 02138, USA, \BGmail}}
\date{\today}

\usepackage{amsmath}
\usepackage{latexsym}
\usepackage{graphicx}

\graphicspath{%
    {converted_graphics/}
    {/}
}

\begin{document}

\maketitle
\vspace{.15in}
\begin{center}{\large\bf for Haim Goldberg (1939$-$2017)}\end{center}
\vspace{.30in}

\begin{abstract}
\noindent Using parametric analysis (curve fitting) we find a persistent temperature bump, coincident with World War \caproman{2} (WW2), in eight independent time series, four land- and four ocean-based. We fit the data with a Gaussian on a quadratic background. Six parameters (constant, linear and quadratic background terms and the amplitude, position and width of the Gaussian) are free to vary. 

The mean fitted Gaussian amplitude is $0.339\pm0.065$\,\degr C, non-zero by $5.2\sigma$ and therefore not accidental. The area is $2.0\pm0.5$\,\degr C\;yr. Temperature recovered to baseline rather quickly. Rather than coincidence, or systematic measuring error synchronized with WW2, we conjecture the bump is due to human activity, including the greatly increased combustion (relative to that era) of fossil and other fuels. 

Background surface temperature behavior, a byproduct of our study but largely independent of the WW2 bump, is far more consequential nowadays. The linear term, $0.747\pm0.023$\,\degr C/century, agrees well with other findings, but the present-day rate of increase, $\approx2.5$\,\degr C/century, is far greater because of the quadratic term. 
\end{abstract}

\clearpage\noindent
Graphs of global surface temperature versus time often show a $0.2-0.3$\,\degr C bump coincident with World War \caproman{2} (WW2). Taken at face value, that is direct evidence of the effect of human activity on global temperature. (There is ample indirect evidence.) The bump could test of climate models if the excess combustion during WW2 could be quantified. Here we show only that the bump is real. It is a persistent feature of the surface temperature record, appearing in eight independent time series, four land- and four ocean-based. A least-squares fit of a Gaussian on a quadratic background to each data set yields values and errors for the constant, linear and quadratic coefficients of the background as well as the magnitude, time of occurrence and duration of the bump. 

\section{Method}

We download twelve NOAA files of temperature anomaly \cite{ERSSTv4}, half land- and half ocean-based. Each set covers the globe in six 30\degr \ zones of latitude and lists monthly surface temperature anomalies and variances from 1880 through 2016, 1644 measurements in all. We arrange and number them in contiguous zones N$-$S, 1$-$6 for land and 7$-$12 for ocean. We fit each zone separately using identical technique (starting values, convergence test $\ldots$) and afterwards, find the weighted mean of seven with acceptable fits. (Data for the four polar zones are too sparse and zone 3 is discussed later.)  

Let $t_i$, $i=1\ldots M$, be the time of the $i^\mathrm{th}$ measurement in years and $T_i$ its temperature anomaly in \degr C. The time range 1880$-$2017 is awkward for a polynomial fit because the coefficients vary by orders of magnitude and have rather obscure meanings: the first coefficient, for instance, is the temperature at year 0. 

Therefore, before fitting, we transform $t$ to a dimensionless variable $u$ ranging from -1 to 1 as $t$ ranges from 1880 to 2017. Because 1 to any power is 1, the resulting polynomial coefficients are all of the same order of magnitude and easier to interpret. For instance, the first coefficient is simply the temperature anomaly at midrange, and $T_M$ at the end of range ($t=t_M, u=1$) is simply the sum of the coefficients. The transformation and its inverse are
\begin{equation}\label{eqn:u}
u\;=\;B(t-A)\quad,\quad t=A+(u/B)
\end{equation} 
where in our case
\begin{equation}\label{eqn:AB}
 A\equiv(t_M+t_1)/2=1948.5\;\hbox{yr}\quad,\quad B\equiv 2/(t_M-t_1)=0.014607\;\hbox{yr$^{-1}$}
\end{equation}
The fit function is
\begin{equation}\label{eqn:F}
F(u)\;\equiv\;p_1+p_2u+p_3u^2+p_6\;e^{\displaystyle{-.5\left((u-p_7)/p_8\right)^2}}
\end{equation}
with $p_6$, $p_7$ and $p_8$ the Gaussian amplitude, mean and $\sigma$ respectively. That form was arrived at after some experimentation (note unused parameter numbers) where we found that a three-term (quadratic) polynomial was the most the data would support. The fit improves with more terms, but their error increases sharply, a sure sign of too many adjustable parameters.

We use the standard Levenberg-Marquardt algorithm to minimize
\begin{equation}\label{eqn:chisq}
\chi^2\;\equiv\;\sum_{i=1}^M\;(T_i-T(u_i;\vec{p}))^2/\sigma_i^2\;=\;
  \sum_{i=1}^M\;w_i\,(T_i-T(u_i;\vec{p}))^2
\end{equation}
with respect to the six parameters $\vec{p}$. The weight $w_i$ of the $i^\mathrm{th}$ point is the inverse of the variance $\sigma_i^2$ given in the NOAA tables. Our form of the algorithm, adapted from Numerical Recipes \cite{nr} allows us (with negligible effort) to hold selected parameters fixed at their initial values. Initial values for the Gaussian are set by hand in an input file and (along with all other program constants) are the same for all 12 data sets to avoid bias. Initial values for the polynomial coefficients are obtained from a preliminary linear least-squares fit without the Gaussian.

The usual criterion for a good fit namely 
\begin{equation}\label{eqn:csqpdf}
\chi^2/(M-6)\approx1
\end{equation}
is too much to hope for here. First, non statistical effects are present, for instance the El Ni\~no Southern Oscillation (ENSO)  in sveral ocean-based zones. Second, Eq.\,\ref{eqn:csqpdf} assumes accurate variances, very unlikely in this complicated a problem. The best we achieve for any zone is $\chi^2/(M-6)\approx4$. Of course, $\chi^2$ still tells us whether one fit is better than another.

After convergence, following \cite{nr}, we call Marquardt once more with $\lambda\doteq0$, whereupon the variance in each fitted parameter equals the corresponding diagonal term of the curvature matrix. We deliberately choose conditions to make the program look fairly hard for the $\chi^2$ minimum: convergence requires $<0.1\%$ improvement in $\chi^2$ and the initial Gaussian amplitude is set at $\approx3\times$ its final value. The program typically converges in $4-6$ passes, except zone\,3 which takes 12 passes to get the wrong answer, as we will see. Fitting all twelve zones takes about 1\,s on an elderly laptop running Intel Visual Fortran.

\section{Results}

\subsection{Preliminary Studies}
Details, tables and graphs can be found in the long version of this report \cite{Gottschalk2017}. Raw land-based (atmospheric) data show much more scatter than ocean-based and for both, scatter decreases from poles to equator. Weights (inverse variances) have the same general pattern, varying by 3$-$4 orders of magnitude month-to-month at the poles. Repeating the final fit with all $w_i\doteq100$ shows that our results are insensitive to the weights, but we did use them for the results below.

Smoothing the raw data by convolution with a Gaussian of $\sigma=24$\,months reveals distinct WW2 bumps in the peri-equatorial ocean-based zones 9 and 10, and hints of bumps in some others. Smoothing can be a useful guide to where to look, but by itself leads to no numbers. 

After fitting, we studied the Gaussian compliance of fit residuals by examining their skewness and kurtosis, both ideally zero, and found some worse than others. ENSO, definitely a systematic effect, seems to be quasi random when averaged over many cycles. 

Finally, we replaced the WW2 Gaussian by a nearly square pulse also having three free parameters (amplitude, start and duration) and found it, too, in all eight zones.

\subsection{Final Results}
\begin{figure}[p] 
  \centering
  \includegraphics[bb=0 0 640 480,width=5.32in,height=4.0in,keepaspectratio]{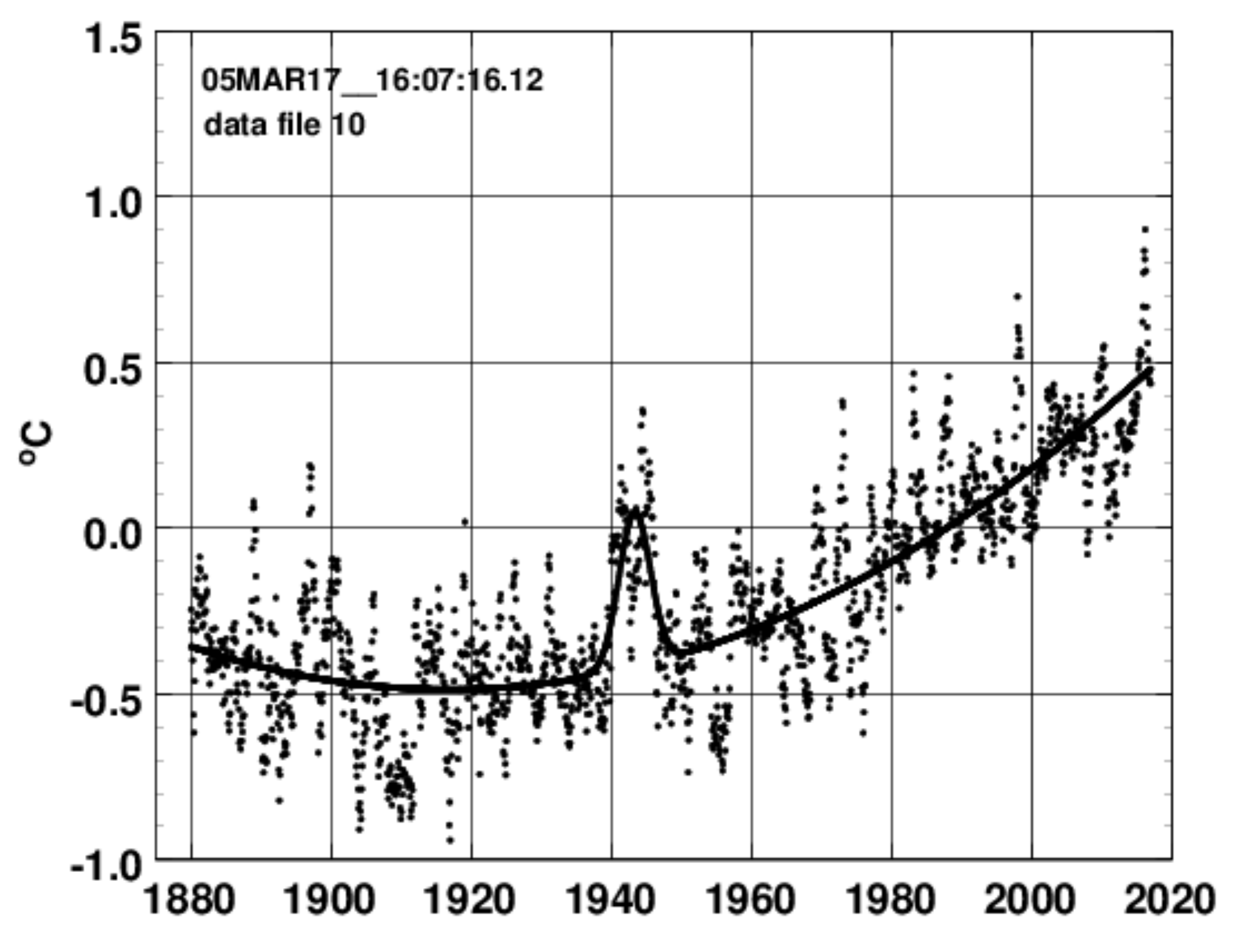}
  \caption{The best fit (zone 10).}
  \label{fig:fitGAU10}
\end{figure}

\begin{figure}[p] 
  \centering
  \includegraphics[bb=0 0 640 480,width=5.32in,height=4.0in,keepaspectratio]{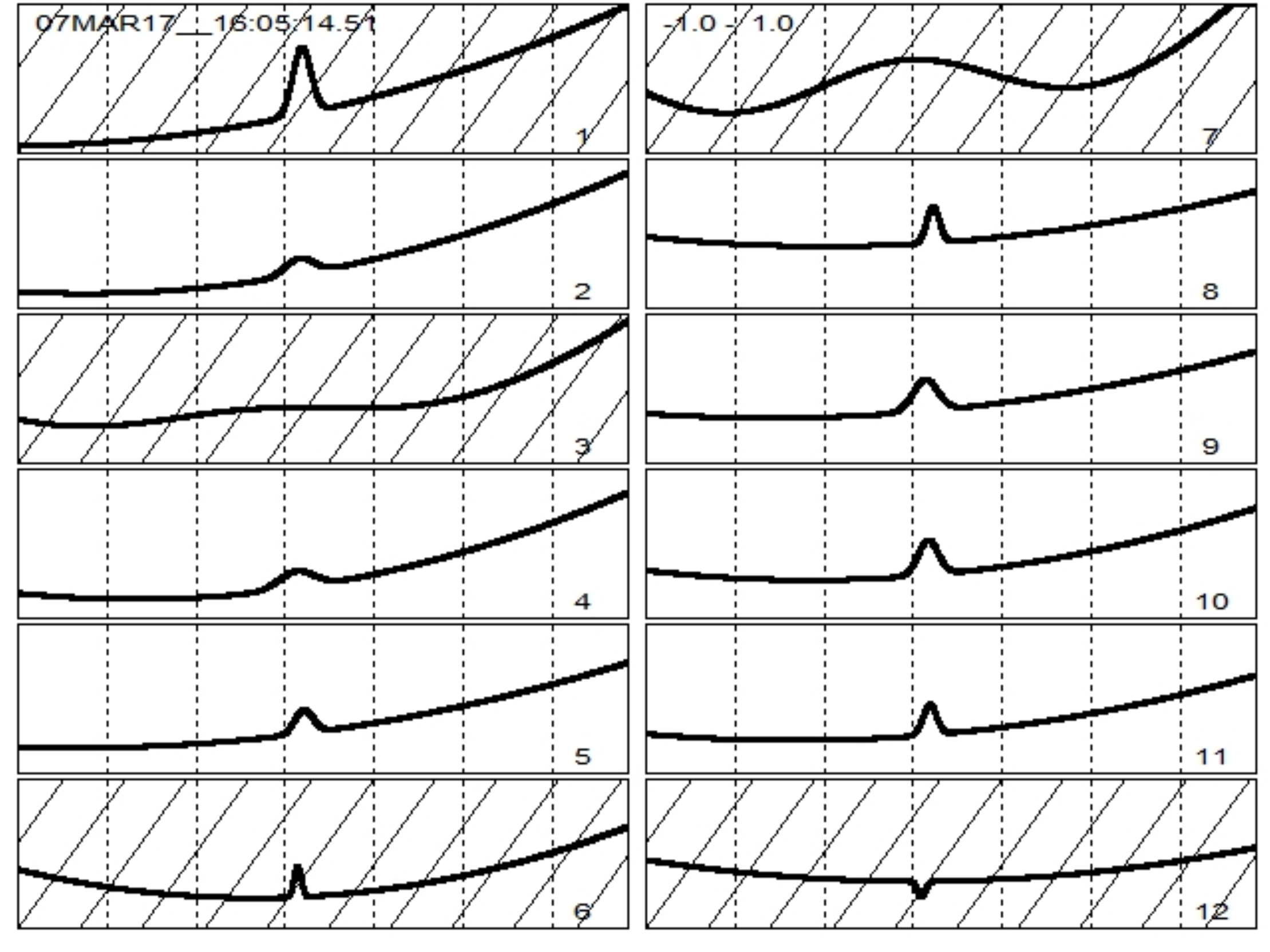}
  \caption{The final cut: fits with a Gaussian on a quadratic background. Cross-hatched zones were dropped.}
  \label{fig:multiyf}
\end{figure}

\fig{fitGAU10} shows the best fit, in zone\,10 (zone\,9 is similar). ENSO is apparent. \fig{multiyf} summarizes all final fits. Left- and right-hand columns are land- and ocean-based respectively, with N at the top. The ordinate range $\pm1$\,\degr C \ is the same for all panels and the abscissa runs from 1880 through 2016 marked every two decades. The five zones that missed the final cut are cross-hatched. 

\begin{figure}[p] 
  \centering
  \includegraphics[bb=0 0 640 480,width=5.32in,height=4.0in,keepaspectratio]{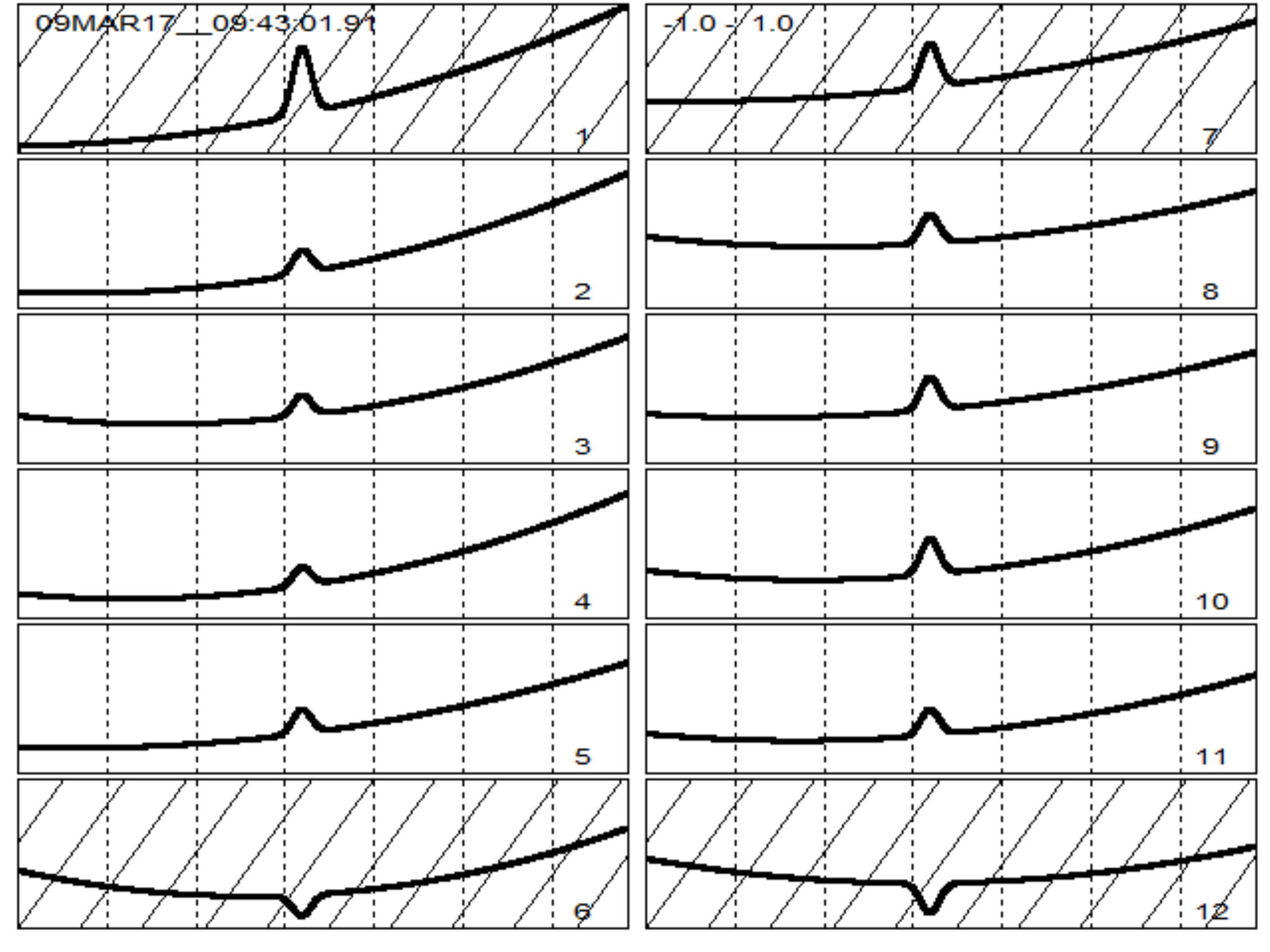}
  \caption{The same fits as \fig{multiyf} with the Gaussian mean (-0.07) and width (0.03) fixed. The amplitude is still free to vary.}
  \label{fig:multiyflock}
\end{figure}

All six parameters are free to vary in all zones. Zone\,3 diverts the three (Gaussian) degrees of freedom so as to improve the background. Jumping to an unsought nearby minimum of $\chi^2$ is a common pathology of nonlinear least-squares fits, and one or another zone does this for every set of starting conditions we tried. \fig{multiyflock}, however, which is the same fit with the Gaussian position and $\sigma$ fixed but the amplitude still free to vary, shows that the bump is also present in zone\,3 with a reasonable amplitude. 

\begin{figure}[p] 
  \centering
  \includegraphics[bb=0 0 640 480,width=5.32in,height=4.0in,keepaspectratio]{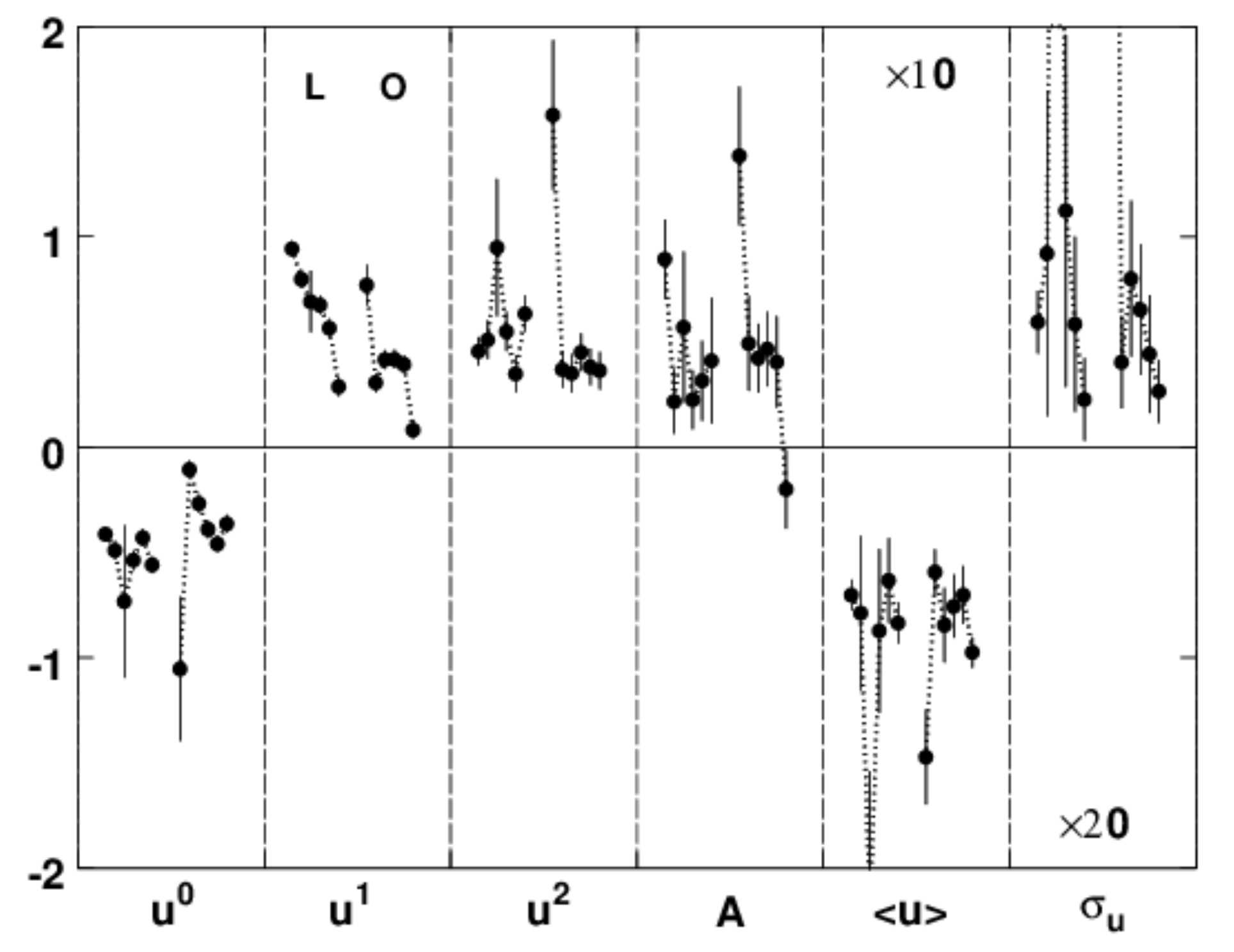}
  \caption{Summary of the three background and three Gaussian parameters, with $1\sigma$ errors.}
  \label{fig:parms}
\end{figure}

Space precludes listing fit parameters by zone here (see \cite{Gottschalk2017}) but \fig{parms} shows the trends. All fits  are shown including dropped zones. Error bars are $1\sigma$; note scale factors and a few off-scale points.
Starting from the left, the difference between L and O constant terms may result from slightly different definitions of temperature `anomaly'. It has no profound consequences. More interesting are the next two blocks showing linear and quadratic terms both larger for L than for O. Both the rate and the acceleration of surface temperature increase seem to be greater for atmospheric than oceanic measurements. Disequilibrium between the atmosphere and the oceans seems to be increasing with time. Relative errors in those numbers are small except for excluded zone\,3.

The Gaussian amplitude block confirms the impression (\fig{multiyf}) that the WW2 bump is smaller in land-based measurements, but the final blocks show that the time of occurrence and duration are, within errors, the same. (We repeat that these were free to vary.) 
  
Table\,\ref{tbl:finalParms} lists seven-zone weighted means of the fit parameters in two ways: in terms of the scaled time variable $u$ used for the fits (left-hand column) and in conventional units (right-hand column). To compute the weighted fit for a given year convert to $u$ using \eqn{u} and insert values from the left-hand column into \eqn{F}, or simply read it off the bold line in \fig{predicted} below. The final two rows of Table\,\ref{tbl:finalParms} list two derived quantities, the areas of the weighted Gaussian and of the weighted square pulse, in excellent agreement. Because each involves both the amplitude and a measure of duration, their relative errors are larger than in the fitted parameters themselves. Since the data do not support a more detailed model of the shape, saying that WW2 was equivalent to a $2.0\pm0.5$\,\degr C\;yr impulse to global surface temperature is probably the most succinct formulation. 

\section{Discussion}
It goes without saying that there are other systematic effects (ENSO being the most dramatic) in the temperature record and there may well be other one-shot events, linked to volcanic eruptions or other phenomena. We simply have not looked for them here.

\fig{predicted} encapsulates the present study, showing fitted results for three land-based and four ocean-based time series as well as their weighted average. We extended it about two decades beyond the present day. A similar figure in \cite{Gottschalk2017} corresponding to the fits in \fig{multiyflock} looks cleaner but leads to the same conclusions.

\subsection{World War \caproman{2} Bump}
According to Table\,\ref{tbl:finalParms} the amplitude $p_6$ of the WW2 bump differs from zero by $5.2\,\sigma$. From standard probability tables the chance of that happening at random is $0.2\times10^{-7}$. If we therefore accept that the effect is real, there remain three possibilities.

First, a systematic measuring error might have come and gone coincident with WW2. That seems superficially plausible; after all, WW2 was a time of global upheaval. However, the systematic aggregation and analysis of surface temperature data cited here did not begin until decades later. Therefore it would have been necessary for a significant fraction of land- and ocean-based stations to change their protocols one way at the beginning of the war and back at the end. The nearest candidate for that is the well known `bucket cooling' effect \cite{Thompson2008}. However, that went in only one direction (after the war ended), it only affected ocean measurements, and ERSST.v4 attempts to correct for it \cite{Huang2015}.

Second, the bump might be pure coincidence: an unrelated climatic event happened to occur at that time, affecting both land- and ocean-based measurements. Logically speaking that argument can never be refuted.

The third possibility is that the WW2 bump was caused by the increase (enormous relative to that era) in the combustion of fossil fuels by the world's armies, navies, air forces and factories, let alone the burning of oilfields, cities and the munitions themselves. Indeed, it would seem inconsistent to believe that burning fossil fuels promotes global warming and, simultaneously, that WW2 had no effect.

\subsection{Background Behavior}
Our study was motivated by curiosity about the WW2 bump, but the background behavior of surface temperature which emerges as a byproduct is of infinitely greater consequence and, mathematically, depends very little on the existence or non-existence of the bump. 

The linear coefficient, $0.747\pm0.024$\;\degr C/century (Table\,\ref{tbl:finalParms}) agrees with $0.735\pm0.068$\;\degr C/century from an independent fit to ERSST.v4 cited by Huang et al. \cite{Huang2015}. Also, if in our analysis we omit the the quadratic term and the WW2 bump and simply fit everything with a straight line, the weighted average is $0.742\pm0.024$\;\degr C/century, so that is all consistent.

A positive quadratic coefficient is required by the mere fact that surface temperature exhibited a minimum around 1910 (as appears in many ocean-based data) or that it was level from (say) 1880$-$1920 and increased thereafter (as is more typical of land-based data). All we have done here is to quantify that as objectively as possible. The positive quadratic coefficient in turn guarantees that the present-day instantaneous rate of change is greater than the long-term average. Unfortunately, looking at \fig{predicted}, it seems to be far greater: about 2.5\,\degr C/century. 

In summary, the good news is that surface temperature reverted to baseline very quickly after WW2. In that era at least, surface temperature was stable with respect to small changes. The bad news is that, because of the quadratic term, the present-day rate of temperature increase is very much larger than the average over the last 137 years.

\section{Acknowledgement}
We thank the Harvard University Physics Department for sustained and generous support. Opinions in this report are ours and not those of the University.


\vspace{1in}
\begin{table}[h]
\setlength{\tabcolsep}{10pt}
\begin{center}
\begin{tabular}{crl}
\multicolumn{1}{c}{quantity}&           
\multicolumn{1}{c}{in terms of $u$}&           
\multicolumn{1}{c}{in ordinary units}\\
\noalign{\vspace{8pt}}           
$p_1$&$-0.3752\pm0.0157\;^\circ$C&$-0.3752\pm0.0157\;^\circ$C\\
$p_2$&$0.5112\pm0.0162\;^\circ$C&$\;\;\;0.7468\pm0.0236\;^\circ$C/century\\
$p_3$&$0.4210\pm0.0334\;^\circ$C&$\;\;\;0.8983\pm0.0713\;^\circ$C/century$^2$\\
\noalign{\vspace{4pt}}
$p_6$&$0.3394\pm0.0652\;^\circ$C&$\;\;\;0.3394\pm0.0652\;^\circ$C\\
$p_7$&$-0.0692\pm0.0062\;\;\;\;\;$&$\;\;\;\;\;1943.76\pm0.43$\;yr\\
$p_8$&$0.0276\pm0.0064\;\;\;\;\;$&$\qquad\;\;\;1.89\pm0.44$\;yr\\
\noalign{\vspace{6pt}}
Gaussian area&&$\qquad\;\;\;1.95\pm0.58\;^\circ$C\;yr\\
square pulse area&&$\qquad\;\;\;1.97\pm0.44\;^\circ$C\;yr
\end{tabular}
\caption{Seven-zone weighted parameters with $1\sigma$ errors, in transformed and ordinary units.}\label{tbl:finalParms}
\end{center}
\end{table}

\clearpage

\clearpage


\begin{figure}[p] 
  \centering
  \includegraphics[bb=0 0 640 480,width=5.32in,height=4.0in,keepaspectratio]{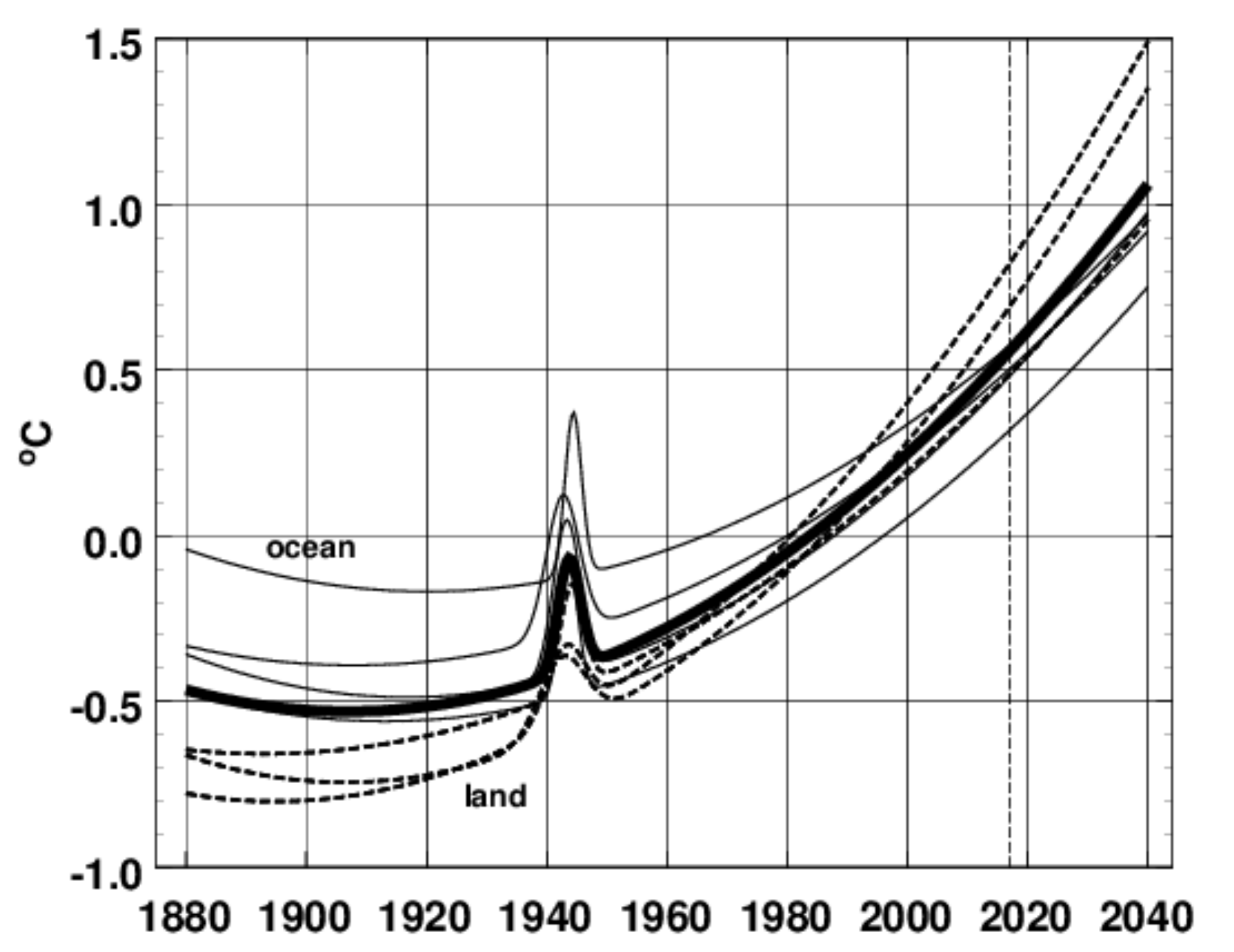}
  \caption{Fitted and extrapolated surface temperatures. Land = dashed line, ocean = light line, weighted mean = bold line.}
  \label{fig:predicted}
\end{figure}

\end{document}